# Data Driven Modeling of Projected Mitigation and Suppressing Strategy Interventions for SARS-COV 2 in Ghana

*Emmanuel de-Graft Johnson Owusu-Ansah(Ph.D)[a,b,c]
Atinuke O. Adebanji[a,b]
Richard K. Avuglah[a]
Eric Nimako-Aidoo[a,b]

[a]Department of Statistics and Actuarial Science, Kwame Nkrumah University of Science and Technology, Kumasi, Ghana.
[b]Laboratory for Interdiscplinary Statistical Applications, Kwame Nkrumah University of Science and Technology, Kumasi, Ghana.
[c]Regional Water, Environmental and Sanitation Center-Kumasi, , Kwame Nkrumah University of Science and Technology, Kumasi, Ghana.
*Corresponding Author: edjowusu-ansah.cos@knust.edu.gh or degraftt@gmail.com/

## Abstract

In the midst of pandemic for respiratory illness, the call for non-pharmaceutical interventions become the highest priority for infectious disease and public health experts, while the race towards vaccine or medical intervention are ongoing. Individuals may modify their behavior and take preventative steps to reduce infection risk in the bid to adhere to the call by government officials and experts. As a result, the existence of relationship between the preliminary and the final transmission rates become feeble. This study evaluates the behavioral changes (mitigation and suppression measures) proposed by public health experts for COVID-19 which had altered human behavior and their day to day lives. The dynamics underlying the mitigation and suppression measures reduces the contacts among citizens and significantly interfere with their physical and social behavior. The results show all the measures have a significant impact on the decline of transmission rate. However, the mitigation measures might prolong the elimination of the transmission which might lead to a severe economic meltdown, yet, a combination of the measures show a possibility of rooting out transmission within 30 days if adhered to in an extreme manner. The result shows a peak period of infection for Ghana ranges from 64[th] day to 74[th] day of infection time period.

Keywords: COVID-19, SARS COV-2, Pandemic Modeling, Epidemics, Ghana.

## 1.0 Introduction

Late December 2019, China reported and confirmed a new pneumonia case caused by a virus called novel corona virus, or SARS-COV-2 (named COVID-19 disease by WHO) within the city of Wuhan in the Hubei province. The virus spread rapidly from 41 confirmed cases to 5,974 within 18 days (10-28 January, 2020), surpassing the SARS pandemic flu. By 4[th] April 2020, COVID-19 had spread to 201 countries/regions across every continent, threatening overwhelming the health-care systems of several countries and threatening to ground global economy. There has been 1,681,954 reported cases with 102,026 deaths as at 10[th] April 2020 (23:58 GMT), and 228,923 recovered. Active cases stood at 810,351 persons worldwide (Worldometers, 2020). China, the epicenter of the pandemic has been surpassed by USA (491,358 cases, 18,316 deaths), Italy (119,827 cases, 14,681 deaths), Spain (119,199 cases, 11,198 deaths), Germany (91,159 cases, 1,275 deaths). In Africa, South Africa is leading with 2,173 cases, 25 deaths, followed by Egypt (2,065 cases, 159 deaths), Algeria (1,914 cases, 293 deaths), Morocco (1,746 cases, 120 deaths), Cameroon (820 cases, 12 deaths), Tunisia (707 cases, 31 deaths), Ivory Coast (566 cases, 5 deaths) and Ghana (566 cases, 8 deaths) as 13[th] April, 2020, 11:48 GMT (Worldometers, 2020).

The severity of transmission appears to be broader than previously thought with the pandemic spreading quickly across the world within a short time and its virulence has been estimated to be high, but, the elderly and those with underlying health conditions experiencing higher mortality rates (Neil M. F., Laydon D., Gemma N-G. et al., 2020). The early stages of the transmission posed challenges for estimating the transmission dynamics of the infection  (Funk S., Ciglenecki I., Tiffany .A, et al., 2017), ). However, recovery over weeks has provided insight into the epidemiological situation and has helped identify whether mitigation and suppression strategies are having desired effects in the populations (Riley S., Fraser C., Donnelly C.A., et al., 2003). The insight elicited can be used to inform prediction of distribution patterns (Viboud C., Sun K., Gaffey R., et al. , 2018), and  estimation of risk in other countries (Cooper B.S., Pitman R.J., Edmunds W.J., et al., 2006). It would also help in the design of alternative non-pharmaceutical interventions (Kucharski A.J., Camacho A., Checchi F, et al., 2015).

While the understanding of the SARS COV-2 and their prevention is not fully known, and the varied prevention strategies for infectious diseases, the world faces a unison challenge from a virus with comparable lethality to H1N1 influenza in 1918.  Two strategies being promoted by health experts based on the evidence from China, Singapore, South Korea and other countries are *Suppression*} and *mitigation* (Neil M. F., Laydon D., Gemma N-G. et al., 2020).

- Suppression: seeks to prevent/reduce secondary transmission.
- Mitigation (Flatten the Curve): at curtailing prevalence rate of the disease (flatten the curve) in order to reduce on health-care providers and facilities and consequently reduce mortality.

This study seeks to assess these two strategies  and their hypothetical impact on COVID-19 transmission   and  management  within  Sub-Saharan Africa (Ghana specifically)given the steady rise of confirmed cases and partial lock-down of the two epicenters of the disease (Greater Accra and Greater Kumasi).

## 2.0 Method

### 2.1 The Growth Models

Pathogens transmission are known to be transmitted in an exponential manner, this is however constrained by natural systems and resulting in the S-shaped curve made possible by the use of ordinary differential equations. The exponential growth of pathogen multiplication is represented with a model without limits. The growth rate is proportional to the population $P$ represented as

$$\frac{dP(t)}{dt} = \alpha P(t) \qquad\qquad 1$$

with

$$P(t) = \beta \times e^{\alpha \times t} \qquad\qquad 2$$

where $\alpha$ is the growth rate and $\beta$ is the initial population of organisms as the population closes to the limit $K$.

To find the growth rate for a short period, natural log is taken on both side resulting as.

$$\ln \beta = \alpha t + \ln P(t) \qquad\qquad 3$$

A simple linear regression is fitted on $\ln \beta$ against $t$ and the gradient $\alpha$ is estimated to be the growth rate.

The logistic model is presented as;

$$\frac{dP(t)}{dt} = \alpha P(t)\left(1 - \frac{P(t)}{K}\right) \qquad\qquad 4$$

solving this for P(t) gives

$$P(t) = \frac{K}{1 + \exp\left(-\alpha(t - \gamma)\right)} \qquad\qquad 5$$

where $\alpha$ is the growth rate, $K$ is the population limit, $\gamma$ specifies the time when the curve reaches $K/2$.

Pathogen transmission is highly dependent on the transmission rate (R0), thus the infectivity of an infectious person is dependent on the latent and infection period of the pathogen therefore, the relationship can be written as

$$R_0 = e^{\alpha\mu - (1/2)\alpha^2\sigma^2} \qquad\qquad 6$$

$$R_0 = \left(1 + \alpha / a\right)\left(1 + \alpha / b\right) \qquad\qquad 7$$

where $R_0$ is the transmission rate, $\alpha$ exponential growth rate, $\mu$ is mean serial interval, $\sigma$ is the standard deviation of the serial interval, $1/a$ is latent period and $1/b$ the infectious period. The expected import of social distancing being promoted by epidemiologists is the reduction of R₀. *Becker,* Niels (2015) proposed a transmission model for social distancing,

$$R = \left[1 - \left(1 - a^2\right) f\right] \times R_0 \qquad\qquad 8$$

where $f$ is the proportion of the population engaging in social distancing to decrease their interpersonal contacts to a fraction $a$ of their normal contacts.

## 2.2 Social Contact Mixing Modeling

The most important aspect of suppressing or mitigating against diseases during pandemic is managing social contact mixing among humans. Data shared through various studies describing social contact mixing were leverage in addition to the online tool to analyse the survey data which captures participant, contact, survey day, household and time use data (see www.socialcontactdata.org). To account for country level social contact rate on a weekly basis, the study make use of weights to account for age and number of social contact within the week (5/7) and weekend (2/7) days. The study adopt weight controlling measure (Eames, K., Tilston, N., White, P et al,. 2010) to limit the influence of single individual social contact rate to 3 to be in line with United Nation's World Population Prospects. The estimation of social contact matrix is represented by (Hens, Ayele, , Goeyvaertset al, 2009) and captured in Lander, Thang, Sebastian et al., 2020.

$$m_{i,j} = \frac{\sum\limits_{t=1}^{T_i} w_{i,t}^d \, y_{i,j,t}}{\sum\limits_{t=1}^{T_i} w_{i,t}^d} \qquad\qquad 9$$

where $w_{i,t}^d$ is the weight for participant $t$ of age $i$ who was surveyed on day type $d \in \left\{weekday, \, weekend\right\}$ and $y_{i,j,t}$ denote the reported number of contacts made by participant $t$ of age $i$ with someone of age $j$. Practically, social contact such as greetings are reciprocal where $m_{i,j}N_i$, is equal to $m_{j,i}N_j$, therefore, to resolve this difference in reciprocity reporting can be imposed by;

$$m_{i,j}^{reciprocal} = \frac{m_{i,j}N_i + m_{j,i}N_j}{2N_i}$$

$$10$$

where $N_i$ and $N_j$ represent the population size in age class $i$ and $j$ respectively (UN, 2019).

The average secondary transmission in age class $i$ by an infected person of age class $j$ into a susceptible population can be represented with the next generation matrix $G$ defined as;

$$G = DMq$$

$$11$$

where $M$ is the contact matrix, $q$ is the proportional factor and $D$ is the mean duration of infectiousness (Held L., Hens, N., ONeil D. et al., 2019; Diekmann O., Heesterbeek J. and Metx J, 1990). The factor measuring proportionality combines several diseases-specific factors that are related to susceptibility and infectiousness indicated in a formulation as;

$$g_{i,j} = D \times m_{i,j} \times s_i \times k_j \times q$$

$$12$$

where $k_j$ represents the infectiousness of age group $j$, $q$ denotes other diseases specific characters and $s_i$ represents the susceptibility of age group $i$. Therefore, the leading right eigenvector of $G$ is proportional to the expected incidence by age and the transmission rate for secondary infection $R_o$ as the dominant Eigen value of $G$ (Chang W, Cheng J, Allaire J, et al, 2017). To evaluate the mitigation and suppression measures, the study focuses on the relative impact of social contact pattern on the rate of second transmission referred to as the social contact hypothesis (Funk S, Socialmixr, 2020) by cancelling out diseases specific characters expressed as '

$$\frac{R_{oa}}{R_{ob}} = \frac{\max\left(eigen\left(D \times M_a \times q\right)\right)}{\max\left(eigen\left(D \times M_b \times q\right)\right)} = \frac{\max\left(eigen\left(M_a \times S \times K\right)\right)}{\max\left(eigen\left(M_b \times S \times K\right)\right)}$$

$$13$$

where subscripts $a$ and $b$ denote different conditions, $S$ and $K$ represent age specific susceptibility and infectiousness respectively (Eurostat, 2019). Hierarchical order is leverage to cater for multiple location contacts such as home, work, transport, leisure, market. Closure of school is simulated by excluding all contacts reported at school before evaluating $m_{i,j}$ and consider increase in social distancing to proportion $p_{socialdis\tan cing}^{t\arg et}$ by accounting for observed social contact at work $M_{work}^{observed}$ and the observed proportion to social distancing $p_{socialdis\tan cing}^{observed}$ given the relation as;

$$M_{work}^{observed} = M_{work}^{all} \times \left(1 - P_{socialdis\tan cing}^{observed}\right) \qquad 13$$

$$M_{work}^{t\arg et} = M_{work}^{all} \times \left(1 - P_{socialdis\tan cing}^{t\arg et}\right) \qquad 14$$

$$M_{work}^{t\arg et} = M_{work}^{observed} \times \left(\frac{1 - P_{socialdis\tan cing}^{t\arg et}}{1 - P_{socialdis\tan cing}^{obsrved}}\right) \qquad 15$$

To combine the effect of social distancing and school closure, the social contact matrix $M$ is calculated as;

$$M = M_{\hom e} + M_{work}^{t\arg et} + \left(M_{school} \times 0\right) + M_{transport} + M_{leisure} + M_{other} \qquad 16$$

The transmission reduction rate from the above formulation is adopted to be used as input in measuring the possible impact for the scenario.

## 3.0 Data Assimilation and Integration

This study also combines different pre-analyzed data from leading epidemiological teams around the world ( Neil M. F., Laydon D., Gemma N-G. et al., (2020); Hermanowicz (2020); Cowling and Leung (2020); Riou and Althaus (2020); Li et al., (2020a); Rabajante (2020); Su et al., (2020); Li et al., (2020b); Shujuan, Jiayue, Minyan et al., 2020; Chang, W., Cheng, J., Allaire, J., Xie, Y. et al., 2017; Geng et al., (2020); Funk, socialmixr, (2020); Lander W., Thang V. H., Sebastian F. et al., (2020)) to form a predictive model to help explain the possible scenarios under the various intervention strategies.

This data is basically gathered from China, UK, Italy, Germany, Spain and South Korea, Japan. Though not much differences exist in terms of factors that facilitate the spreading of the SARS COV-2 in these regions. The common factors among these countries include the weather, settlement patterns and a robust health-care facilities comparatively far better than those found in Sub-Saharan Africa. However, some values such as the transmission rate or the basic reproduction number is very important as it gives information to the average number of secondary infections which is vital for prediction in any setting irrespective of prevailing local conditions .

Different basic reproduction numbers ($R_0$) or rate for secondary infection for COVID-19 have been reported by different studies from these countries (Table 1). These values ranges from 1.4 to 7.0, confirming the virulent nature and the uncertainty surrounding its infectious rate in different studies. COVID-19 come from the same family of viruses as SARS and display similarities, but the   high transmission rates poses the greatest threat to mankind over a century.

**Table 1: Summary of Estimated Basic Reproduction Number (Ro) for COVID-19**

| Epidemic | Reproduction Number | References |
|---|---|---|
| SARS | 2.0-5.0 | Wallinga and Teunis (2004) |
| Influenza | 2.0-3.0 | Mills et al. (2004) |
| Ebola | 1.5-2.5 | Althaus (2014) |
| COVID-19 | 1.4-2.5 | WHO(1.4-2.5); Hermanowicz (2020)(2.4-2.5); Cowling and Leung (2020)(2.2); Riou and Althaus (2020)(2.2); Li et al. (2020a)(2.2); Rabajante (2020)(2.0); Su et al. (2020)(2.24-3.58); Li et al., (2020 )( 2.2-3.1); Geng et al. (2020)(2.38-2.72) |
|  | 2.5-3.0 | Zhou et al. (2020) (2.8-3.9); Su et al. (2020) (2.24-3.58); Geng et al. (2020) (2.38-2.72); Xiong and Yan (2020) (2.7); Li et al., (2020b) (2.2-3.1); Wu et al. (2020a) (2.68) |
|  | 3.0-3.5 | Zhou et al. (2020) (2.8-3.9); Su et al. (2020) (2.24-3.58); Li et al., (2020c) (2.2-3.1); Liu et al. (2020) (3.28); Cao et al. (2020b) (3.24); Read et al. (2020) (3.11); Cao et al. (2020a) (3.24 |
|  | 3.5-4.0 | Zhou et al. (2020) (2.8-3.9); Su et al. (2020) (2.24-3.58); Zhang et al. (2020) (3.6) |
|  | 4.0-7.0 | Shen et al. (2002) (4.71); Sanche et al. (2020) (4.7-6.6) |

Credit: Xiuli, Geoffery, Wang, Qin, Xiang, Ziheng and Li, 2020

Recent study in UK  (Shujuan, Jiayue, Minyan et al., 2020.), estimated the basic reproduction number used for intervention measurement from 2.0 to 2.6, such transmission rate is a reflection of the planned structures, family size and community engagements which exist in the country, it is expected that, the values of Ghana might be higher, however, expected to fall within the range of previous studies as shown in Table 1 .

As reported, incubation period is assumed to be 5.1 days (Linton N.M., Kobayashi T., Yang Y., et al., 2020; Li Q., Guan X., Wu P., et al., 2020). Infectiousness is 12 hours for pre-symptomatic and symptomatic and 4.6 days for asymptomatic (Riou J, Althaus CL. Pattern, 2020).\\

Symptomatic individuals are 10-50\% more infectious than asymptomatic individuals and a uniform distribution is fitted to quantify for the uncertainty in variations of infection, additionally, infectiousness seeded in each country shows an exponential rate. Thus, doubling every 5 days (Neil M. F., Laydon D., Gemma N-G. et a.l, 2020). All other input values are shown on Table 2,

**Table 2: Parametric Values**

| Parameter | Values | References |
|---|---|---|
| Incubation Period | 5.1 days | Linton NM, Kobayashi T, Yang Y, et al, 2020; Li Q, Guan X, Wu P, et al., 2020 |
| Pre-Symptomatic Infectiousness | 12 hours-4.6 days | Riou J, Althaus CL. Pattern, 2020 |
| Infectious doubling | 4-5 days | https://ourworldindata.org/coronavirus#our-data-sources<br>Neil M F, Laydon D, Gemma N-G etl al, 2020 |
| Growth rate | Varied | Estimated fro Equation |
| Transmission rate (R0) | 2.2 -2.6 | Shujuan, Jiayue, Minyan et al., 2020; Natalie, Tetsuro, Yichi et al., 2020; Refer to Table 1 |
| Latent Period | 2.25-3.95 days | Shujuan, Jiayue, Minyan et al., 2020; Natalie, Tetsuro, Yichi et al., 2020. |
| Infectious Period | 7.10 – 7.28 days | Shujuan, Jiayue, Minyan et al., 2020; Natalie, Tetsuro, Yichi et al., 2020. |

**3.1 Intervention Strategies**

The model analyzes the impact of mitigation and suppression on reducing the rate of transmission and demand on health-care facilities by spreading the demand over a long period of time. Data on mitigation and suppression were gathered from publicly available data of the ongoing COVID -19 epidemic. Data for the various interventions are as shown on Table 3.

Table 3: Mitigation and Suppression Measures

| Intervention | Description | References |
|---|---|---|
| Closure of Schools and Universities | Closure of all schools, Household contact increases by 50%. Community contact increases by 25%. 10% Reduction in transmission rate | Neil M F, Laydon D, Gemma N-G etl al, 2020; Lander, Thang, Sebastian et al., 2020 |
| Case Isolation | Symptomatic person stays at home for 7 days. Non-house hold contact reduced by 70%, household contacts unchanged. Assuming 50% of household adhere to this. | Neil M F, Laydon D, Gemma N-G etl al, 2020 |
| Voluntary Home Quarantine | Following identification of a symptomatic case in the household, all household members remain at home for 14 days. Household contact rates double during this quarantine period, contacts in the community reduce by 75%. Assume 50% of household comply with the policy. | |
| Social distancing of entire Population | All households reduce contact outside household, school or workplace by 75%. School contact rates unchanged, workplace contact rates reduced by 25%. Household contact rates assumed to increase by 25%. | |
| Partial Lockdown | All households reduce contact outside household school or workplace by 80% and 90% of households in affected areas adhere to it. | |

## 4.0 Analysis and Results

The study considered 3 scenarios, first is a 'no intervention' (Business as usual) situation, where the virus is allowed to move through the population seamlessly. The second scenario is putting in place mitigation measures, which includes voluntary home quarantine, case isolation, social distancing of some extent (partial lock-down) to most susceptible group. Lastly, the suppression measures which combines case isolation, social distancing of the entire population, and either household quarantine or school and university closure as reported by Stephanie Soucheray (2020).

## 4.1 Results from Social Contact Mixing

The use of online Socrates (socialcontact.org) tool to study the effect of school closure and teleworking on transmission rates within the African countries were studied. The respondents' age were categorized into 4 classes, 0-18 years, 19-45 years, 45-60 years and 60+ years. Holiday period data was excluded before contact rate was estimated. The reference proportion of telework was fixed at 5\% to present relative increase in Telework and transmission dynamics were captured from 10\% to 70\% telework with school closure. The impact on transmission was made to reflect the vulnerability of the elderly and was fixed at 0.5, 1, 1.5, 2.0 respectively. The study found out, these measures impact the transmission rate by 10% reduction, therefore closure of school and increase telework will have more contact in the community but reduction of transfer of pathogens among children which is a source of pathogen transfer to adults. This result was incorporated into the other measures to reflect the situation in Ghana, since the president ordered for school and universities closures before all other mitigation measures were laid out.

### *Data fitting*

Fitting the COVID-19 data in Ghana (cumulative data points as at 12[th] April, 2020) to the exponential model, the model recorded a growth rate of 0.151794 (0.116759 to 0.186829) resulting in a transmission rate of 2.31 to 4.08. These values gave more insight into the possible modeling of future epidemics and how the measures taken by government will impact on the spread. This result (Table 4, see \ref{table4}) helps in the further modeling procedure of examining the peak time and the peak number of COVID-19 cases. Ghana reported its first case on 12[th] March, 2020, and had its first compulsory quarantine of all foreign arrivals on 21[st] March, 2020, making all cases up to this date had a possible community asymptomatic transmission.

Table 4: Results of Data Fitting for Parameters from COVID-19 Ghana.

| Rate | Value | Std Error | 95% CI |
|---|---|---|---|
| **Rate** | 0.151794 | 0.016796 | 0.116759 to 0.186829 |
| **Initial** | 9.470796 | 2.940384 | 3.337337 to 15.60425 |
| | COVARIANCE | | |
| | | Initial | Rate |
| | Initial | 0.015020 | -0.000085 |
| | Rate | -0.000085 | 0.000000 |

## 4.2 Results for COVID-19 Projections

Results in Table 5(see \ref{table5}) show estimated descriptive values for projected possible number of transmissions by the model for the three different scenarios. It also takes into consideration quantifying of the uncertainty in the projections by estimating for both minimum and maximum possible projections of transmission for 30 days and 60 days of COVID-19 projections for Ghana

## 4.3 Projection of Possible Confirmed Cases

Ghana will continue to have increasing confirmed cases of COVID-19 for the next two weeks, these cases will be strictly increasing until the peak time (Figure 1). The increment projections are as a result of the continuous comprehensive mass testing and surveillance in addition to the contact tracing currently underway. As indicated in Figure 1, the confirmed cases might rise up steadily to 1000 cases in not too distant future by 21[st] April, 2020, this projection is to give indication to the respond team to plan ahead in the management of cases that might come out as a result of the testing currently taking place.

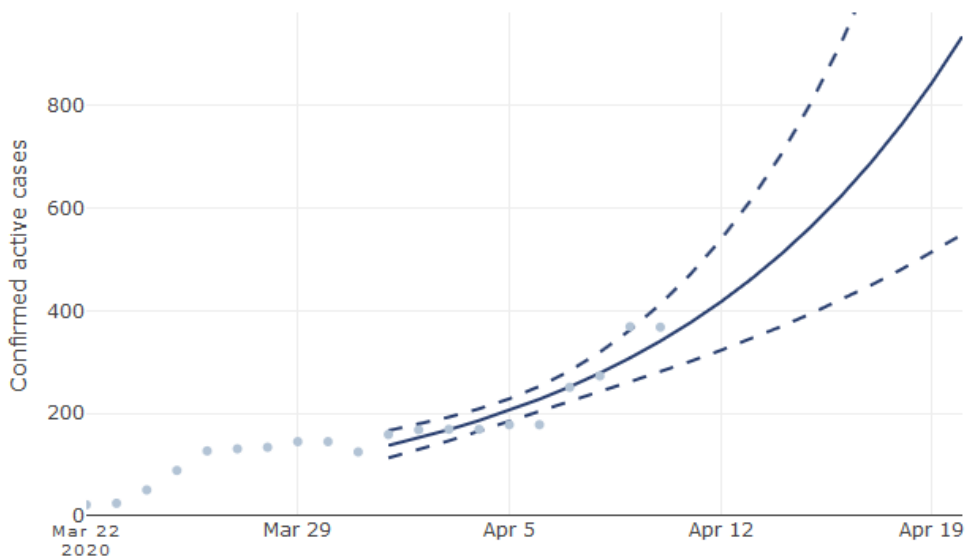

Figure 1: Projection of Possible COVID-19 Cases in Ghana simulated from the number of confirmed cases.

### 4.4. Projections for No Intervention Scenario (Business as usual)

The "Business as usual" scenario projection (Figure 2, Table 5) transmission ranges from 6,774 to 55,436 transmission cases with modal value of 6, 845 cases projected for 30 days in Ghana if mass testing of affected areas are carried out. The 60 days projections was estimated between 138,440 and 15,064,182, thus, approximately 0.4% to 53.8\% of the population in a free fall 60 days transmission, the modal projected estimation for 60-days stands 229,751. The projected transmission number could be higher for Ghana in both the 30-days and 60 days projections if authorities continues to confirm more horizontal (community transmission) cases which are not previously known, or if the vertical cases (imported) patients are found to have spent days within the communities as asymptomatic carriers of the virus. "Quote: 'The assumption is that, the total number of COVID-19 cases is not known. It is however certain that the total number of COVID-19 cases is higher than the number of confirmed cases. This is due to limited testing, without widespread testing for COVID-19 we can neither know how the pandemic is spreading nor appropriately respond to it. The best preliminary research data will need to be revised as the pandemic progress and new cases confirmed. This is so because a new reported case on any specified day do not necessarily represent a new case on that particular day' unquote (ourworldindata, 2020)".

### 4.4.1 Mitigation Measures

The mitigation measures (voluntary home quarantine, case isolation, Partial lock-down) show a significant reduction in transmission. Each measure ensures a substantial decline in cases if strictly adhered to, using the modal value, case isolation brings the projected cases to 1,631.79 and with possible range of 1,615.72 to 11,698.38. Voluntary quarantine has a modal value of 2,023.82 and ranges from 2,003.71 to 14,737.83 possible projected cases and that of partial lockdown has a modal value of 1,194.16 and ranges from 1,182.53 to 8,359.97 (Table 5, see \ref{table5}). The predictions exceeds critical bed capacity, and this is a necessary measure that needs to be in place as long as the epidemic is in force.

Among the mitigation measures, case isolation and partial lock-down are the most effective methods to flatten the curve. The modal projected transmission keeps declining with each mitigation adopted, however, it should be noted, in these scenarios the closure of school and universities are inclusive, since these measures have already been implemented by the government as the first strategy before other mitigations measures were considered, the impact of the mitigation measures is attributed to the decline in contact within the younger generation.

### 4.4.2 Suppression Measures

The introduced suppression measures (Table 5) had the most impact in decreasing the transmission projected numbers for each scenario It, should be emphasized that, the 30-day suppression measures was every effective in reducing the projected transmission number.. A combination of Case Isolation, Voluntary Quarantine and Social Distancing of entire population and partial lock-down projected transmission is the most effective way of managing the epidemic. However, these measures should be adhered to in an extreme manner in order to achieve the results. When examining these mitigation and suppression measures, the study assumes a force adherence of 30 to 90 days or longer in order to effectively achieve the desirable results. Overall, the relative effectiveness of the measures is sensitive to the combination of scenarios chosen to be in force in order to control the epidemics.

### 4.5 Peak Time Analysis

The peak period for maximum infection is expected to be within the 69th day after the first infection with uncertainty quantification putting the period anywhere between 63rd day to 74th day after infection. The total number of infections if nothing is done could reach 75.43\% (Figure 3 and Figure 4) of the population from the projections. Given the time to peak infection, the model conclusion is hinged on a possible extension of social distancing longer than the prescribed 14-day partial lock-down. As shown on Figure 1, the peak probable cumulative infection number could reach 100,000 plus infection by the peak time if mass testing and other measures are not effectively enforced. However, COVID-19 has been observed to display different pattern of reproduction in malaria endemic countries, this could be reduced as a result of herd immunity due to the use of hydroxychloroquine in treatment of COVID 19 (myjoyonline, 2020).

It is imperative that the suppression measures be enforced by government to achieve a reproduction rate less than 1. This is necessary given the dire state of the health-care facility in Ghana and its gross inadequacy to provide the required services needed in the event of a major outbreak. The uncertainties of the effectiveness of mitigation measures notwithstanding, it provides the surest way to prevent an escalation of COVID-19 cases in Ghana when coupled with  individual measures such as cleaning of hands with soaps under running water, avoidance of crowded spaces, avoidance of traditional hand greetings and the use      of      hand      sanitizers      as      well      as      staying      home.

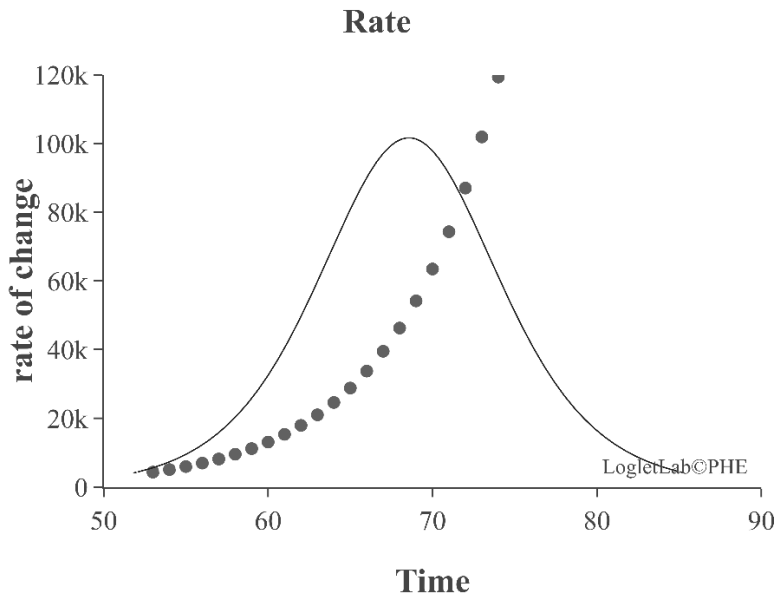

Figure 2: Time to Peak and Maximum peak graph simulated from possible infection number for Ghana.

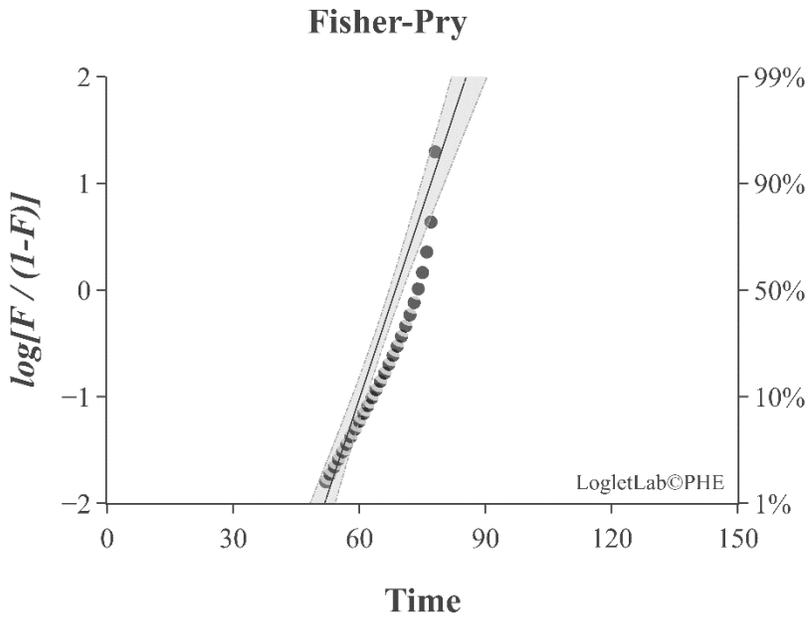

Figure 3: Projection of COVID-19 cumulative infection in Ghana simulated from possible infection number.

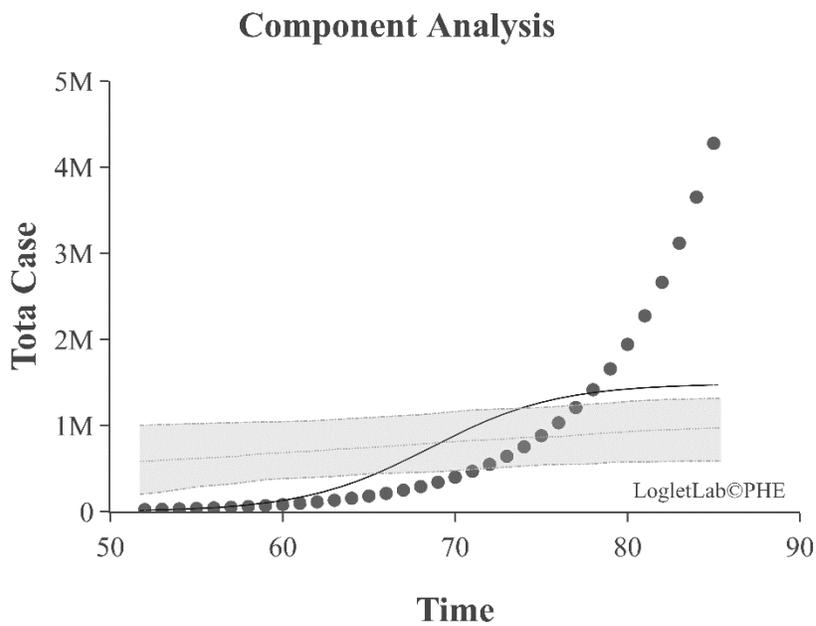

Figure 4: Projection of COVID-19 infection in Ghana simulated from possible infection number.

**Table 5: Comparison of Interventions**

| Intervention Strategies | Projections | Mean | Mode | Min | Max | Standard Deviation |
|---|---|---|---|---|---|---|
| **Do Nothing** | **30 Days** | 23,149.24 | 6,845.99 | 6,774.34 | 55,435.49 | 13,571.24 |
| | 60 Days | 3,529,725.18 | **229,751.81** | 224,959.26 | 15,064,182.07 | 3,811,193.66 |
| **Case Isolation** | 30 Days | 5,093.12 | 1,631.79 | 1,615.72 | 11,698.38 | 2,822.36 |
| **Voluntary Home Quarantine** | 30 Days | 6,381.73 | 2,023.82 | 2,003.71 | 14,737.83 | 3,562.92 |
| **Social distancing of entire population** | 30 Days | 2,042.46 | 685.45 | 678.94 | 4,576.08 | 1,093.16 |
| **Partial Lockdown** | 30 Days | 3,669.83 | 1,194.16 | 1,182.53 | 8,359.97 | 2,010.52 |
| **Case Isolation + Voluntary Home Quarantine** | 30 Days | 296.81 | 239.32 | 238.82 | 363.51 | 35.94 |
| **Voluntary Home Quarantine + Partial Lockdown** | 30 Days | 286.12 | 233.67 | 233.21 | 346.49 | 32.66 |
| **Case Isolation + Voluntary Home Quarantine + Partial Lockdown** | 30 Days | 212.437 | 205.956 | 205.892 | 219.120 | 3.819 |
| **Case Isolation + Voluntary Home Quarantine + Social distancing of entire population** | 30 Days | 210.892 | 205.348 | 205.293 | 216.592 | 3.262 |

**5.0 Discussions**

The present study on COVID-19 transmission were explicitly estimates projected using publicly available data. The outcome shows effectiveness of mitigation and suppression measures to flatten the transmission curve, even though the 30-day projections seems encouraging for declining projected transmission number, an ideal situation is a 3-4 months suppression measures in order to control the epidemic (Stephanie Soucheray, 2020). The estimated projections for combination of measures and within time frame is consistent with the practice in China, Japan, Hong Kong and Taiwan which have recorded cases below that predicted by experts (Xiuli Liu, Geoffrey Hewings, Shouyang Wang , Minghui Qin, Xin Xiang, Shan Zheng, Xuefeng, 2020). It should be pointed out that , the estimations are in line with projections by experts in UK and goes on to confirm the possibility of adopting these measures in order to defeat the virulent transmission of COVID-19 (Neil M F, Laydon D, Gemma N-G etl al, 2020).

The relative effectiveness of the measures from countries which have experienced the impact of COVID-19 cannot be underestimated, due to the fact that the implementation of parts or combinations of these measures have successfully resulted in achieving some sort of positive impact up to date, New Zealand, China, Singapore, Italy and New York (USA) are examples. To large extent, population wide social distancing within the entire country would have had the largest impact to control the virus transmission and bring the basic reproduction number less than 1, thereby rapidly reducing the case incidence within the population. Even though, the results show a possible control of COVID-19 within 30 days, to avoid re-insurgence due to reinfection, the measures need to stay longer to eliminate the transmission completely. As argued out by Neil M F, Laydon D, Gemma N-G et al. (2020) Ghana as a developing country cannot as a feasible policy sustain prolonged suppression measures due to the economic impact it might have and a possible recession halting the rebound of the national economy, a 3-4 months suppression with intermittent relaxation when improvement is realized is the viable option whiles waiting for a pharmaceutical intervention such as vaccine is developed.\\\\

Several limitation to the present study exist. Firstly, the study relies on published information on the ongoing crisis, though, no country or study has been able to put out the full picture or understands the transmission completely. This information are based on the decisions made by the organizations or government agencies that put out the information. Secondary, the study did not take into consideration the effect of high temperature had on the transmission level, as Ghana is a tropical country with high temperatures known to be far higher than what experts say the virus could survive under (Fauci, 2020). Finally, the study uses growth rate which have been recorded at the early stage of transmission subject to intrinsic and extrinsic health factors in the various countries where the cases were imported from. Whiles we acknowledge these limitations, we believe the current study throws more light on possible impacts of the interventions and feasibility of projections given data availability.

**Supplementary Materials:** All important information are found in the Tables

**Conflicts of Interest:** The authors declare no conflict of interest.

 

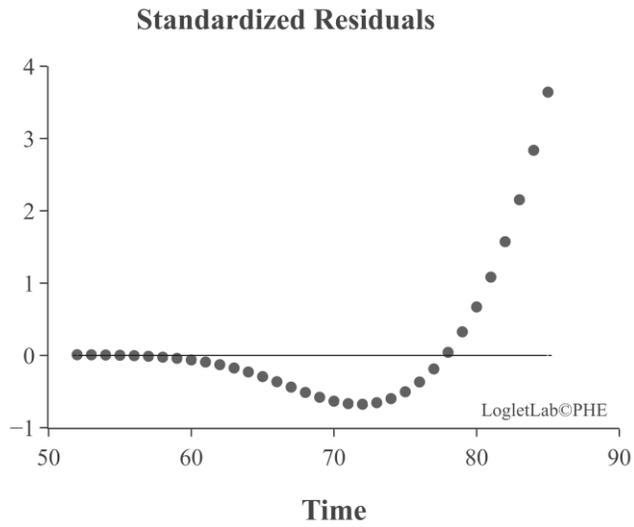

Figure 5: Standardize residual of projection model.

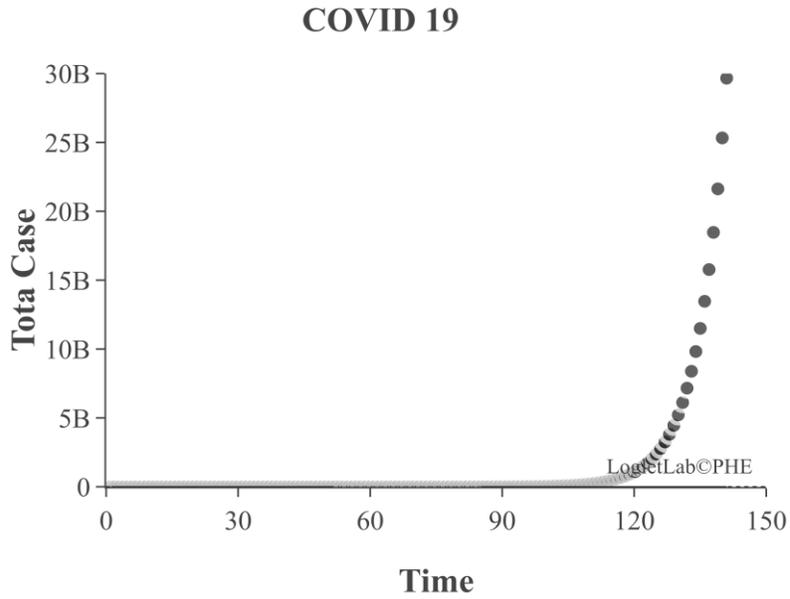

Figure 6: Projections of COVID-19 without intervention beyond 120 days.